\documentstyle[preprint,aps,epsf,psfig]{revtex}
\begin{document}
\title{Lamellae alignment by shear flow in a model of a diblock
copolymer}
\author{Fran\protect\c{c}ois Drolet$^{1}$ and Jorge Vi\~nals$^{1,2}$}
\address{$^{1}$ Supercomputer Computations Research Institute,
Florida State University, Tallahassee, Florida 32306-4130, USA
$^{2}$ Department of Chemical Engineering,
FAMU-FSU College of Engineering, Tallahassee, Florida 32310-6046, USA
}

\date{\today}

\maketitle

\begin{abstract}

A mesoscopic model of a diblock copolymer is used to study the
stability of a lamellar structure under a uniform shear
flow. We first obtain the nonlinear lamellar
solutions under both steady and oscillatory 
shear flows. Regions of existence of these solutions are determined
as a function of the parameters of the model and of the flow. Finally,
we address the stability of the lamellar solution against long
wavelength perturbations.

\end{abstract}

\section{Introduction}
\label{sec:introduction}

We study a mesoscopic model of a block copolymer to describe the 
re-orientation of a lamellar
structure by an imposed uniform shear flow that is either 
constant or periodic in
time. This is a first step towards understanding
known phenomenology pertaining to the response of the
block copolymer microstructure to shear flows near the isotropic to 
lamellar transition
\cite{re:zchen97}. The model that we use is based on the free 
energy of the diblock copolymer obtained by Leibler
\cite{re:leibler80}, and later by Ohta and Kawasaki 
\cite{re:ohta86}, to which an
advection term is added to incorporate the effect of 
the externally applied shear
flow. We identify spatially periodic solutions
that correspond to a lamellar structure, and determine their stability
against a number of long wavelength perturbations.

Modulated phases are ubiquitous in physical and chemical systems
\cite{re:seul95}. They generally result from the competition between short
and long range forces. Additional
symmetries of the system (e.g., translational or rotational invariance)
often lead in practice to rich textures, especially in systems of
extent that is large compared with the characteristic wavelength of the
modulation. Modulated phases often have interesting macroscopic behavior,
and exhibit a complex response to externally 
applied forces. While it is possible to devise approximate
constitutive laws to describe the macroscopic response of such phases,
it is often necessary to explicitly address their evolution at the mesoscopic
scale, and to determine how microstructure evolution influences 
the macroscopic response.

We focus here on the lamellar phase observed in diblock copolymers below
the order-disorder transition
\cite{re:leibler80,re:ohta86,re:bates90}. Diblock copolymers are
formed by two distinct sequences of monomers, A and B, that are
mutually incompatible but chemically linked. At sufficiently low
temperatures, species A and B would segregate to form macroscopic 
domains, but the chemical
bonding between the two leads to a modulated phase instead.
The detailed equilibrium microstructure depends on the relative 
molecular weight of the chains 
\cite{re:leibler80,re:ohta86,re:matsen94,re:laradji97}
and has been studied in detail within a mean field 
approximation \cite{re:netz97,re:villain-guillot98}. 

We follow in this paper the approach of
Leibler who introduced an order parameter field $\psi({\mathbf r})$ that
describes the local number density difference of monomers A and B. The
order parameter is defined to be zero above the order-disorder
transition, and is finite and nonuniform below. Leibler's analysis was
restricted to the weak-segregation limit (close to the order-disorder
transition) within which the thickness of the interface separating the
A-rich from the A-poor regions is of the order of the wavelength
of the microstructure. Later, Ohta and Kawasaki extended Leibler's
free energy to the strong segregation range, and showed the importance
of long ranged effective interactions that arise from the 
connectivity of the polymer chains. We
use this latter free energy as the driving force for the re-orientation
dynamics, allowing also for passive advection of the
order parameter by an imposed shear flow. The model
studied is similar to that considered by Fredrickson
\cite{re:fredrickson94}, except that we neglect thermal
fluctuations and assume that both phases have the same viscosity. 

The stability of a lamellar structure to secondary instabilities has
already been addressed in the literature, although in the absence of shear
flow \cite{re:kodama96,re:shiwa97}. In fact, the similarity between 
the equations governing the motion of the
lamellae and the Swift-Hohenberg model of Rayleigh-B\'enard convection 
\cite{re:swift77,re:greenside85,re:cross93} gives rise to a common 
phenomenology \cite{re:shiwa97}. The lamellar
structure is found to
be stable only within a range of wavenumbers. At higher wavenumbers it
undergoes an Eckhaus instability which generally results in a decrease
of wavenumber, whereas for wavenumbers below that range the structure
undergoes a zig-zag instability. In this paper, we extend these
stability results to explicitly include fluid advection by the imposed
shear. We find that the stability boundaries are modified with
respect to the zero velocity case in a way that depends not only
on the amplitude of the shear, but also on the orientation of the
lamellae relative to the flow. Of course, the latter dependence is
absent in earlier treatments that neglected advection.

Our results are a first step towards understanding the complex
re-orientation
phenomenology that has been observed experimentally
\cite{re:zchen97}. In this initial analysis, we introduce a number 
of restrictive
assumptions that we plan to relax in future work. First, our
calculations are primarily two dimensional and thus can only address
the so-called parallel and transverse orientations. Second, and more
importantly, we neglect thermal fluctuations and any viscosity contrast 
between the two phases,
elements that have been argued to be important in determining the
main qualitative features of the
re-orientation process. We also neglect flow induced by the
lamellae themselves in response to the applied shear. These secondary
flows could become important for the late stage coarsening of the 
lamellar structure. Finally, we have
confined our study to locating the boundaries of several 
secondary instabilities of the lamellar
structure, but have not addressed the evolution following the
instabilities, nor the coarsening of the resulting textured pattern 
\cite{re:elder92,re:cross95a}. However, our results concerning the
periodic base solution under flow, and its stability against long
wavelength perturbations are the prerequisite building blocks of 
a more general theory. 

\section{Mesoscopic model equations}
\label{sec:menm}

Following Leibler \cite{re:leibler80}, we introduce
an order parameter field, $\psi(\mathbf{r})$, function of the local
density difference of monomers A and B. For a block copolymer
with equal length sub-chains, the order parameter is
$\psi ({\mathbf{r}})= \frac{\rho_A({\mathbf{r}})- \rho_B({\mathbf{r}})}
 {2 \rho_{0}}$, where $\rho_X, X=A,B$ is the density of
monomer $X$, and $\rho_{0}$ is the total density, 
assumed constant (incompressibility condition).
A mean field free energy ${\cal F}\left[ \psi(\mathbf{r}) \right]$ was
derived by Leibler for a monodisperse diblock copolymer
melt \cite{re:leibler80}, and later by Ohta and Kawasaki
\cite{re:ohta86}. In units of $k_{B}T$, where $k_{B}$ is Boltzmann's
constant and $T$ the temperature, the free energy is comprised of two terms,
${\cal F}/(\rho_{0}k_{B}T) = {\cal F}_s + {\cal F}_l$.
The term ${\cal F}_s$ incorporates local monomer interactions,
\begin{displaymath}
{\cal F}_s = \int  \mbox{d {\bf{r}}} 
 \left[ \frac{\kappa}{2} |\nabla \psi|^2 - \frac{\tau}{2} \psi^2
+ \frac{u}{4} \psi^4 \right],  
\end{displaymath}
and is formally identical to the Ginzburg-Landau free energy commonly
used to describe phase separation in a binary fluid mixture 
\cite{re:gunton83}.
The parameters $\kappa, \tau$ and $B$ can be approximately
related to the polymerization
index $N$, Kuhn's statistical length $b$ and the Flory-Huggins
parameter $\chi$ through the relations 
$ \kappa = \frac{b^2}{3}$, $\tau=\frac{2 \chi N - 7.2}{N}$ 
and $ B = \frac{144}{N^2 b^2}$ \cite{re:kodama96}.

Long range interactions
arising from the covalent bond connecting the two sub-chains 
are contained in ${\cal F}_l$,
\begin{displaymath}
{\cal F}_l = \frac{B}{2} \int \! \!\int  {\mathrm d} {\mathbf r}  
  \, {\mathrm d} {\mathbf r'} G({\mathbf r-r'}) \psi({\mathbf r})
\psi({\mathbf r'}) 
\end{displaymath}
where the kernel $G({\mathbf r-r'})$ is the infinite space Green's
function of the Laplacian operator
$\nabla^2 G(\mathbf{r-r'}) = - \delta(\mathbf{r-r'})$.
The nonlocal interactions arising from the connectivity of the 
chains lead to a thermodynamic
equilibrium state with a nonuniform density. In our case
of equal length sub-chains, the equilibrium
configuration is a periodic lamellar structure,
with a characteristic wavelength of the order of 100 \AA ~ for a
typical system.

Given this mean field free energy, a phenomenological set of equations
that govern the temporal relaxation of equilibrium thermal fluctuations 
of $\psi(\mathbf{r})$ and of fluid velocity $\mathbf{v}$ has
been derived  close to the order-disorder transition 
\cite{re:fredrickson88,re:helfand89,re:fredrickson94}. 
A similar phenomenological description can be used below the
order-disorder transition under the
assumption that the local relaxation of the order parameter field at the
mesoscopic scale is still driven by minimization of the same free energy 
\cite{re:gurtin96,re:anderson98}. Under this assumption,
$\psi$ obeys the time-dependent Ginzburg-Landau equation,
\begin{equation}
\label{eq:mod4}
\frac {\partial \psi}{\partial t} + {\mathbf v} \cdot {\bf{\nabla}} \psi
= M \nabla^2 \frac{\delta {\cal F}}{\delta \psi},
\end{equation}
where $M$ is a phenomenological mobility coefficient,
and $\delta / \delta \psi$ stands for functional differentiation with
respect to $\psi$. 
Equation (\ref{eq:mod4}) includes the effect of advection
by a local velocity field {\bf v}, which satisfies an extended Navier-Stokes 
equation 
\begin{equation}
\label{eq:ns}
\frac{\partial \mathbf{v}}{\partial t} + ({\mathbf v} \cdot {\mathbf \nabla})
{\mathbf v} = \nu \nabla^2 {\mathbf{v}} - \frac{{\mathbf \nabla} p}{\rho}
 + \frac{\delta 
{\cal F}}{\delta \psi}  \frac{{\mathbf \nabla} \psi}{\rho},
\end{equation}
where $\nu$ is the kinematic viscosity of the fluid,
assumed constant and 
independent of $\psi$, $p$ is the fluid pressure, and 
appropriate boundary conditions for both $\psi$ and {\bf v} must be 
introduced. The last term
on the right-hand side of Eq. (\ref{eq:ns}) is required to ensure that
there cannot be free energy reduction by pure advection of $\psi$
\cite{re:gurtin96}. This term is sometimes referred to as osmotic
stress, and it leads to the creation of rotational flow by
curved lamellae that is directed towards their local center of curvature.

We focus on a layer of block copolymer,
unbounded in the $x$ and $y$ directions, and
being uniformly sheared along the $z$ direction (Fig. 
\ref{fi:schem}).
The layer is confined between the stationary $z=0$ plane,
and the plane $z=d$ which is uniformly displaced parallel
to itself with a velocity  $v_{\mathrm{plane}} = s \, d$ in the case
of a steady shear, and $v_{\mathrm{plane}} = \gamma \, d \, \omega \, 
\cos(\omega t)$ in the case of an oscillatory shear.
$s$ is the dimensional shear rate in the steady case, and $\gamma$ is 
the dimensionless strain amplitude in the case of an oscillatory shear
of angular frequency $\omega$.

The general problem defined by Eqs. (\ref{eq:mod4}) and (\ref{eq:ns}) 
can be considerably simplified by noting that under typical
experimental conditions inertia is negligible ($\omega d^2/\nu \ll
1$). Furthermore, we will
neglect in this paper the term
$(\delta {\cal F}/\delta \psi) {\mathbf \nabla} \psi/\rho$ in Eq. 
(\ref{eq:ns}). Under these conditions,
Eq. (\ref{eq:ns}) admits a simple solution that satisfies the
specified conditions at the moving plane: 
${\mathbf v}= s \, z \, \hat{i}$ 
for a steady shear, and 
${\mathbf v}= \gamma \, d \, \omega \cos(\omega t) \, 
z \, \hat{i}$ for an oscillatory shear, where $\hat{i}$ is the unit
vector in the $x$ direction. 
Therefore, the problem reduces to a 
single governing equation for the order parameter field 
$\psi$ (Eq. (\ref{eq:mod4})) under a prescribed advection velocity 
${\mathbf v}$. As discussed in the introduction, previous theoretical
work on the formation and
stability of lamellar structures further neglected advection of $\psi$
in Eq. (\ref{eq:mod4}).
The results presented in this paper are free of this restriction.

Since the base state to be considered is comprised of spatially
uniform lamellae advected by the shear flow, it is convenient
to introduce a new frame of reference in which the velocity
vanishes. Define a new system of non mutually orthogonal coordinates 
$(x_{1},x_{2},x_{3})$ by $ x_{1} = x - a(t) z, x_{2} = y$ 
and $ x_{3} = z$. The dimensionless quantity $a(t) = s 
\, t$ for a steady shear,
and $a(t)=\gamma \sin(\omega t)$ for an oscillatory shear.
All the calculations reported in this paper, both analytical and
numerical, have been performed in this new frame of reference.
Analytical calculations consider an unbounded geometry in the $x_{1}$ and
$x_{2}$ directions and periodic boundary conditions along the $x_{3}$
direction, whereas the numerical computations
have been conducted in a two dimensional, square domain on the 
$(x_{1},x_{3})$ plane 
and consider periodic boundary conditions along both $x_{1}$ and 
$x_{3}$. Note that both frame of references coincide at $t=0$, and at
equal successive intervals of one half the period of the shear in the case
of oscillatory shear.

Dimensionless variables are introduced by defining
a scale of length by $\sqrt{\kappa/\tau}$, a scale of time by
$\kappa /M\tau^2$, and an order parameter
scale by $\sqrt{\tau/u}$. 
In the transformed frame of reference and in dimensionless
variables, Eq. (\ref{eq:mod4}) reads,
\begin{equation}
\label{eq:main}
\frac{\partial \psi}{\partial t}  =
\nabla'^2 (\psi + \psi^3 - \nabla'^2 \psi) - \frac{B \kappa}{\tau^{2}} \psi,
\end{equation}
with
\begin{displaymath}
  \nabla'^2 = \left[ 1 + a^2(t) \right] 
\frac{\partial^2}{\partial x_{1}^2} - 2 a(t) 
\frac{\partial^2}{\partial x_{1} \partial x_{3}} +  
\frac{\partial^2}{\partial x_{3}^2}
+  \frac{\partial^2}{\partial x_{2}^2}.
\end{displaymath} 
There is only one dimensionless group remaining $B \kappa /\tau^2$,
which will be simply denoted by $B$ in what follows.

We will first show in Sec. \ref{sec:weak} that below (but close to) the 
order-disorder transition point (in the weak-segregation limit), Eq. 
(\ref{eq:main}) admits periodic solutions. Their stability 
against infinitesimal long wavelength perturbations is the subject of 
Sec. \ref{sec:si}.

\section{Lamellar solution in the weak-segregation limit}
\label{sec:weak}

In the absence of shear ($a(t) = 0$) the uniform solution of Eq. 
(\ref{eq:main}), $\psi = 0$, loses stability at the order-disorder 
transition. In a mean field approximation the transition
occurs at $B_{c} = 1/4$. This is a supercritical bifurcation with a 
critical wavenumber $q_{c} = \sqrt{1/ 2}$. 
Near threshold $( 0 \le \epsilon = (B_{c} - B)/2B_{c} \ll 1)$
there exist periodic stationary solutions of the form
\begin{equation}
\label{eq:solu}
\psi({\mathbf r}) = 2 A \cos ( {\mathbf q \cdot r})
+ A_1 \cos (3  {\mathbf q \cdot r})  + \ldots ,
\end{equation}
with $ A^{2}  = \frac{q^2 - q^4 - B}{3 q^2} \sim O(\epsilon)$, and 
$A_1$ of higher order in $\epsilon$. 
This solution only exists for a range of wavenumbers $q$ such that
$\sigma(q^2) = q^2 - q^4 - B \geq 0$.

\begin{figure}[ht]
{\psfig{figure=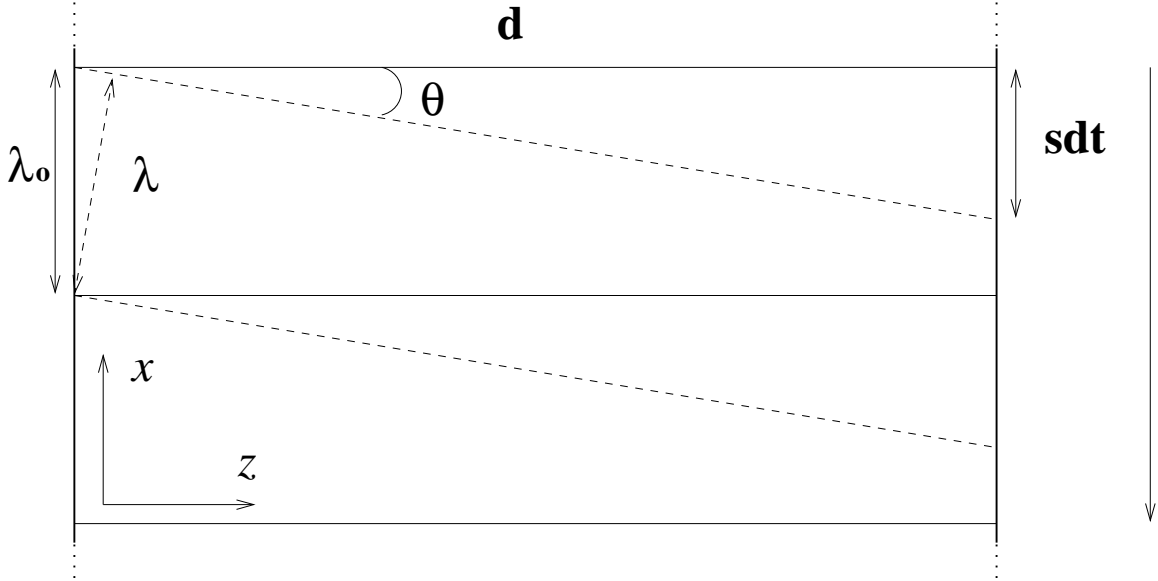,width=6in}}
\caption{Schematic representation of the configuration studied. We also
show a schematic of the distortion of a lamellar pattern under
uniform shear flow. The imposed velocity field is along the $x$
direction. The velocity is specified at the $z = d$ boundary, and 
vanishes at $z=0$.
The lamellae in this graph are transverse to the flow at $t=0$ (solid
lines). At a later time (dotted lines) the lamellae are at an angle
with respect to the flow, and the wavelength has
changed accordingly.}
\label{fi:schem}
\end{figure}

For nonzero shear, we seek solutions of Eq. (\ref{eq:main})
of the form of Eq. (\ref{eq:solu}), with ${\mathbf r} = (x_{1},
x_{2}, x_{3})$  expressed
in the sheared frame basis set $ \{ {\mathbf e}_{1} = \hat{i}, 
{\mathbf e}_{2} = \hat{j},{\mathbf e}_{3} = a(t) \hat{i} + \hat{k} \}$. 
Wavevectors are expressed in the
reciprocal basis set $ \{ {\mathbf g}_{1} = \hat{i} -
a(t) \hat{k}, {\mathbf g}_{2}=\hat{j}, {\mathbf g}_{3} =
\hat{k} \}$. Therefore we keep
the same functional form as for nonzero shear, but
allow a time-dependent amplitude $A(t)$. Note that the components of
the wavevector ${\mathbf q}$ are assumed to be independent of time
and given by $q_1=q_x(t=0), q_2=q_y(t=0)$ and $q_3=q_z(t=0)$ respectively.
The wavevector itself depends on time through the time
dependence of the reciprocal basis set.
Such a solution corresponds to a spatially uniform lamellar 
structure with a time-dependent wavevector that adiabatically follows
the imposed shear in the laboratory frame (see Fig. \ref{fi:schem}).
Inserting Eq. (\ref{eq:solu}) into Eq. (\ref{eq:main}), we find to
order $\epsilon^{3/2}$ ($\sigma$ is itself of order $\epsilon$), 
\begin{equation}
\label{eq:dif}
\nonumber \frac{d A}{d t} = \sigma \left[ q^{2}(t) \right] A - 3 q^{2}(t) A^3,
\end{equation}
with $q^{2}(t) = q_{1}^{2} +\left( a(t) q_{1} - q_{3} \right)^{2} +
q_{2}^{2}$, and $\sigma(q^{2}) = q^{2} - q^{4} - B$.
This nonlinear equation with time-dependent coefficients can be solved
exactly in the two cases of steady and oscillatory shear flow.

In the case of a steady shear $a(t)= s t$. We find,
\begin{equation}
\label{eq:solss}
A(t) = \left\{ \frac{e^{2 H(t)}}{A(0)^2} + 6 e^{2 H(t)} 
\int_0^t dt'\, e^{-2 H(t')} q^{2}(t') \right\}^{-1/2},
\end{equation}
with 
\begin{equation}
H(t)= ( q_0^4 + B - q_0^2 ) \; t + (1 - 2 q_0^2 ) s q_1 q_3 
 \; t^2 + \frac{2 q_1^2 s^2 (2 q_3^2 + q_0^2)}{3} \; t^3
- q_1^3 q_3 s^3 \; t^4 + \frac{q_1^4 s^4}{5} \; t^5.
\end{equation}
The constant quantity $q_0 = \sqrt{q_{1}^{2}+q_{2}^{2}+q_{3}^{2}}$
is the initial wavenumber, and $A(0)$ is the initial amplitude.
For the special case $q_1=0$,  $A(t)$  simply relaxes to 
its equilibrium value in the absence of shear 
$ A^{2}  = \frac{q_0^2 - q_0^4 - B}{3 q_0^2}$. 
This corresponds to an initial orientation of the structure which has no 
component transverse to the flow. For any other initial orientation,
the shear induces changes in the lamellar spacing 
in the laboratory frame of reference (Fig. \ref{fi:schem}).
As a result, the amplitude $A(t)$ decreases and approaches zero at long times.
Hence, the structure melts and reforms with
a different orientation which we cannot predict on the basis of our
single mode analysis. The emerging structure presumably
results from the amplification
of thermal fluctuations near the point at which the amplitude $A(t)$
vanishes, and they have been neglected in our treatment. Thermal
fluctuation effects have been accounted for by others 
\cite{re:cates89,re:fredrickson94}.

For an oscillatory shear $a(t)= \gamma \sin(\omega t)$. We first
examine the stability of the uniform solution $\psi = 0$ against small
perturbations. Linearization of Eq. (\ref{eq:dif}) leads to,
\begin{equation}
\label{eq:floquet1d}
\frac{d A(t)}{dt} = \sigma \left[ q^{2}(t) \right] A(t),
\end{equation}
with $\sigma(t+T) = \sigma(t)$ and $T =  2 \pi / \omega$.
Equation (\ref{eq:floquet1d}) constitutes a one-dimensional Floquet
problem. The solution $A=0$ is unstable when 
\begin{equation}
\bar{\sigma} = \int_{0}^{T} \sigma(t) dt > 0.
\end{equation}
The resulting neutral stability curve is given by,
\begin{equation}
\label{eq:floq_int}
B = q_0^2 - q_0^4 - \frac{3 q_1^4 \gamma^4}{8} 
-\frac{(2 q_0^2 + 4 q_3^2 - 1) \gamma^2 q_1^2}{2}.
\end{equation}
Instability modes can be conveniently classified by considering the
relative orientation of the lamellae at $t=0$ and the shear
direction. We define a parallel orientation, $q_{3} \neq 0, q_{1}
= q_{2} = 0$; a perpendicular orientation, $q_{2} \neq 0, q_{1} =
q_{3} = 0$; and a transverse orientation, $q_{1} \neq 0,
q_{2}=q_{3} = 0$. The following instability points are
identified depending on the orientation of the critical wavevector:
a transverse mode with
\begin{equation}
B_{c} = \frac{1}{2} \frac{ (2 + \gamma^{2} )^{2}}{8 + 8 \gamma^{2} + 3
\gamma^{4}}, ~~ q_{1c} = \sqrt{ \frac{4+2 \gamma^{2}}{8 + 8 \gamma^{2}
+ 3 \gamma^{4}}},
\end{equation}
a mixed parallel-perpendicular mode with
\begin{equation}
B_{c} = \frac{1}{4}, ~~ q_{1c} = 0, ~~~ 2 q_{2c}^{2} + 2 q_{3c}^{2} = 1,
\end{equation}
and a mixed parallel-transverse mode defined by
\begin{equation}
B_{c} = \frac{1}{4} \frac{7 \gamma^{2} + 16}{15 \gamma^{2} + 16},
~ q_{2c} = 0, ~ q_{1c} = 2 \sqrt{ \frac{1}{15 \gamma^{2} + 16}}, ~ q_{3c} =
\sqrt{ \frac{3 \gamma^{2} + 8}{30 \gamma^{2} + 32}}.
\end{equation}
Note that the threshold corresponding to perturbations of
wavevectors that do not
have a projection along the transverse direction are not affected by
the shear. Furthermore, neither the stability boundaries nor the values
of the critical wavenumbers depend on the angular frequency $\omega$.

\begin{figure}
\centerline{\psfig{figure=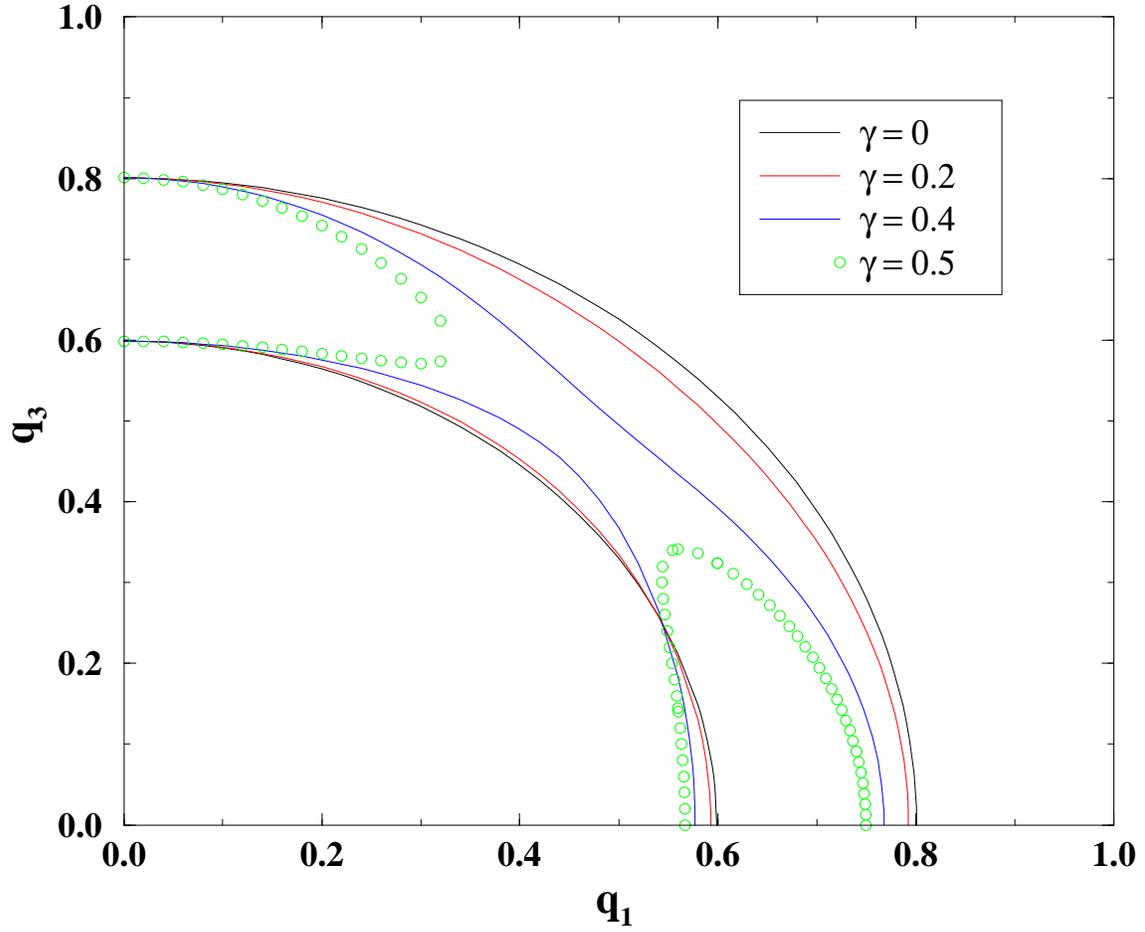,width=6in}}
\caption{Neutral stability curves at $\epsilon = 0.04$ as a function
of the wavenumber of the perturbation at $t=0$ $(q_{1}, q_{3})$ for the values
of the dimensionless strain rate $\gamma$ indicated. The inner regions bounded
by the various curves represent the regions in which the solution
$\psi = 0$ is linearly unstable.}
\label{fi:neutral}
\end{figure}

In what follows, we consider mainly two dimensional solutions in the plane
$q_{2} = 0$ (transverse and parallel orientations) to make contact with
two dimensional numerical calculations. 
As an example, Fig. \ref{fi:neutral} shows the
neutral stability curve in the $(q_{1},q_{3})$ plane for mixed
parallel-transverse modes at $\epsilon = 0.04$, and for several values
of the dimensionless strain amplitude $\gamma$. Recall that $q_{1} =
q_{x}(t=0)$ and $q_{3}=q_{z}(t=0)$ define the initial orientation of
the lamellae. The figure shows that
the shear does not modify the neutral stability curve in the vicinity
of $q_{1} = 0$ (parallel orientation), whereas the curve is shifted
near $q_{3} = 0$ (transverse orientation). Large changes are observed for
oblique wavevectors, including the complete suppression of
the instability at sufficiently large values of the strain amplitude.

Above threshold, Eq. (\ref{eq:dif}) can be solved to yield the
time-dependent amplitude of the lamellar structure under oscillatory
shear. We find,  
\begin{equation}
\label{eq:sol}
A(t) = \left\{ \frac{e^{2 (I(t) - c_4 - c_5)}}{A(0)^2} + 6 e^{2 I(t)} 
\int_0^t dt'\, e^{-2 I(t')} q^{2}(t') \right\}^{-1/2}.
\end{equation}
The function $I(t)$ is given by 
\begin{equation}
I(t) = c_1 t + c_2 \sin (2 \omega t) + c_3 \sin (4 \omega t) 
+ c_4 \cos (\omega t)  +  c_5  \cos^3 (\omega t),
\end{equation}
with
\begin{itemize}
    \item[] $c_1 = \left[ \frac{3 q_1^4 \gamma^4}{8} +\frac{(2 
q_0^2 + 4 q_3^2 - 1) \gamma^2 q_1^2}{2} + q_0^4  + B -
q_0^2 \right]$ \\
    \item[] $c_2 = -\left[ \frac{q_1^4 \gamma^4 + ( 2 q_0^2 + 4
q_3^2 - 1 ) \gamma^2 q_1^2}{4 \omega} \right]$, ~~~~ 
     $c_3 = \frac{q_1^4 \gamma^4}{32 \omega}$ \\
    \item[] $c_4 = \left[ \frac{(4 q_0^2  - 2 ) \gamma q_1 q_3 + 
         4 \gamma^3 q_1^3 q_3}{\omega} \right]$,~ and  ~~
     $c_5 = - \frac{4 \gamma^3 q_1^3 q_3}{3 \omega}$. \\
\end{itemize}

We note that the stability condition Eq. (\ref{eq:floq_int}) is
equivalent to $c_1 =  0$.
Hence, the asymptotic behavior of $A(t)$ at long times changes 
qualitatively depending on the sign of $c_1$. For $c_1 > 0$,
$\lim_{t  \rightarrow \infty} e^{-2I} = 0 $, so that the integral in 
Eq. (\ref{eq:sol}) tends to a finite constant. Since the prefactor 
$e^{2I}$ diverges exponentially, $A(t)$ decays to zero. 
If, on the other hand, $c_{1} < 0$, $A(t)$ becomes periodic at
long times. To prove this statement, we first rewrite the second
term inside the curly brackets in Eq. (\ref{eq:sol}) as
\begin{equation}
\label{eq:asym}
{\cal I}(t) = 6 e^{2 f(t)} \int_0^t dt' \, e^{-2 c_1 (t'-t) - 2 f(t')}
q^{2}(t'), 
\end{equation}
with $f(t')= c_2 \sin (2 \omega t') + c_3 \sin (4 \omega t')
+ c_4 \cos (\omega t')  +  c_5  \cos^3 (\omega t')$. 
Since both $f(t')$ and $q^{2}(t')$ are periodic with period  $T=2 \pi/\omega$,
we can decompose  Eq. (\ref{eq:asym}) into
\begin{equation}
\label{eq:asym2}
{\cal I}(t)= 6 e^{2 f(t)} \left[ \sum_{j=1}^{n} e^{2 j c_1 T} \int_0^T dt' \,  
 e^{-2 c_1 t' - 2 f(t'+t)} q^{2}(t'+t) + e^{2 c_1 t} \int_0^{t-nT}  dt' \, 
 e^{-2 c_1 t' - 2 f(t')} q^{2}(t') \right],
\end{equation}
where  $n$ is an integer such that $0 < t - nT < T$. In the limit of
large $t$ and with $c_1$ negative,
the last term on the right-hand side of Eq. (\ref{eq:asym2}) 
vanishes while the sum $ \sum_{j=1}^{n} e^{2 j c_1 T}$ converges 
to $1/(e^{-2 c_1 T}-1)$. Combining Eqs. (\ref{eq:sol}) and (\ref{eq:asym2}) 
yields an asymptotically periodic solution for $A(t)$,
\begin{equation}
\label{eq:sol2}
 \nonumber
 A(t) =  \left[\frac{6 e^{2 f(t)}}{e^{-2 c_1 T}-1} \int_0^T dt' \,  
 e^{-2 c_1 t' - 2 f(t'+t)} q^{2}(t'+t) \right]^{-1/2}.
\end{equation}

The condition $c_1=0$ can
also be understood in terms of a critical strain amplitude $\gamma_{c}$
above which an existing lamellar structure of a given orientation at
$t=0$ will melt (i.e., $A(t)$ will decay to zero at long times) 
The value of $\gamma_c$ that corresponds to $c_{1}=0$ is
given by,   
\begin{equation}
\label{eq:gamc}
\gamma_c=[(-b+ \sqrt{b^2 - 4 d c})/2d\,]^{1/2},
\end{equation}
with $b= (2 q_0^2 + 4 q_3^2 - 1) q_1^2/2, \; c = q_0^4 + B - q_0^2$,
and $d=3 q_1^4/8$. Note again that the
critical strain amplitude is independent of the angular frequency $\omega$.

\begin{figure}[ht]
\centerline{\psfig{figure=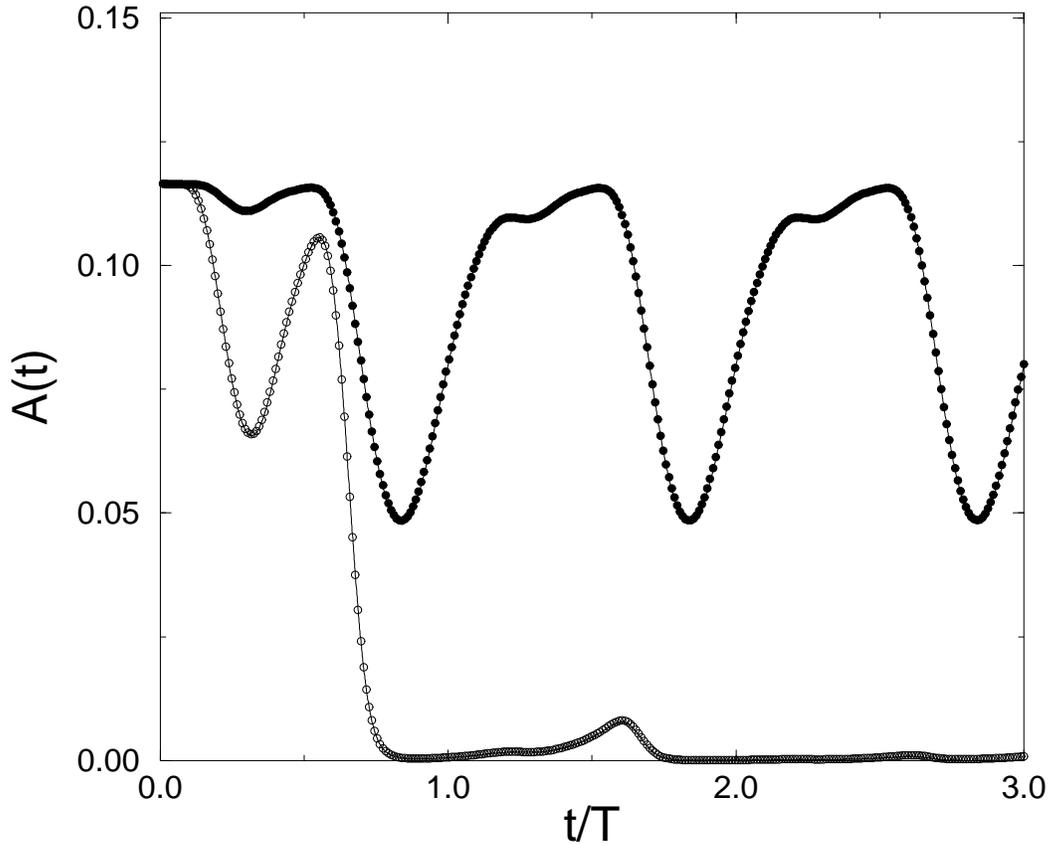,width=6.0in}}
\caption{Temporal evolution of the amplitude $A(t)$ of the single mode
solution given in Eq. (\ref{eq:sol}) for an oscillatory shear
(solid lines), along with the corresponding numerical solution of the
full model for $\epsilon = 0.04$, $\omega=0.02$ and $\gamma = 0.5$ ($\bullet$),
and $\gamma = 0.75 $ ($\circ$). Time has been scaled by the period of 
the shear $T=2 \pi / \omega$. For $\gamma = 0.75$, the initial
periodic pattern is unstable against uniform melting and the amplitude 
$A(t)$ decreases to zero. On the other hand, for $\gamma = 0.5$ a 
spatially periodic solution is stable, and after a transient, $A(t)$
becomes a periodic function of time (Eq. (\ref{eq:sol2})).}
\label{fig:avst}
\end{figure}

In order to test the approximations involved in
Eq. (\ref{eq:solu}), namely that the wavevector ${\mathbf q}$ adiabatically
follows the flow, and the single mode truncation for small $\epsilon$,
we have undertaken a numerical solution of the model equation in a
two dimensional, square geometry (see the Appendix for the details of
the numerical method). As a first example, we consider
an oscillatory shear of angular frequency $\omega=0.02$ imposed on a lamellar
structure of initial wavevector $(q_1,q_3)=(0.687,0.098)$.
The critical strain amplitude for this initial orientation is 
$\gamma_c=0.695$.
Figure \ref{fig:avst} shows the temporal evolution of $A(t)$ for two 
values of $\gamma$, one larger and one smaller than $\gamma_c$. 
The solid lines are the predictions of Eq. (\ref{eq:sol}), and the
symbols are the results of the numerical calculation.
The agreement in both cases is excellent.

\section{Secondary instabilities of the lamellar pattern}
\label{sec:si}

In order to address the stability of the lamellar pattern, we next consider
long wavelength perturbations of the base state with
wavevector ${\mathbf Q} = (Q_{1},
Q_{2}, Q_{3})$ such that its components are also constant in the sheared frame 
of reference. Close to threshold, perturbations evolve in a slow
time scale compared to the inverse frequency of the shear. We
therefore assume that the wavenumber of any long wave perturbation
would adiabatically follow the imposed flow. Specifically, we consider
a solution of the form,
\begin{equation}
\label{eq:full}
\psi({\mathbf r},t) = [A(t) + \delta A_{+} e^{i {\mathbf Q  \cdot r}} +
\delta A_{-} 
 e^{-i {\mathbf Q  \cdot r}}] e^{i {\mathbf q  \cdot r}}  + {\rm c.c.}
\end{equation}
where $A(t)$ is the nonlinear solution obtained in 
Section \ref{sec:weak}.
Substituting Eq. (\ref{eq:full}) into Eq. (\ref{eq:main}), and linearizing 
with respect to the amplitudes $\delta A_+$ 
and $\delta A_-$, we find,
\begin{equation}
\label{eq:lsa}
\frac{\partial}{\partial t} \left[ \begin{array}{c} {\delta A_{+}} \\ 
 {\delta A_{-}} \end{array}  \right]  =    L(t) 
\left[ \begin{array}{c}  \delta A_+ \\ \delta A_- \end{array} \right],
\end{equation}
with
\begin{displaymath}
 L(t) = \left[ \begin{array}{cc}
- l_{+} -  l_{+}^2 - B + 6 A(t)^2 l_{+} & 3 A(t)^2 l_{+} \\
3 A(t)^2 l_{-} & - l_{-} - l_{-}^{2} - B + 6 A(t)^{2} l_{-}
\end{array} \right],
\end{displaymath}
and $l_\pm = - ({\mathbf q} \pm {\mathbf Q})^{2} =
-(1 + a(t)^2) (q_1 \pm Q_1)^2 + 2 a(t) (q_1 \pm Q_1)(q_3 \pm Q_3)
-(q_{2} \pm Q_{2})^{2}-(q_3 \pm Q_3)^2$.
In general, the matrix elements $L_{ij}$ are complicated functions of time, 
and we have not attempted to solve Eq. (\ref{eq:lsa}) analytically. 
For $\gamma<\gamma_c$ the operator $L$ contains terms that are both
periodic in time and decaying transients. At long enough times,
$A(t)$ is given by Eq. (\ref{eq:sol2}), and the linear system Eq. 
(\ref{eq:lsa}) has periodic 
coefficients. Hence, it reduces to a two dimensional Floquet problem 
for the amplitudes $\delta A_{+}$ and $\delta A_{-}$ \cite{re:iooss90}.

\begin{figure}[ht]
{\psfig{figure=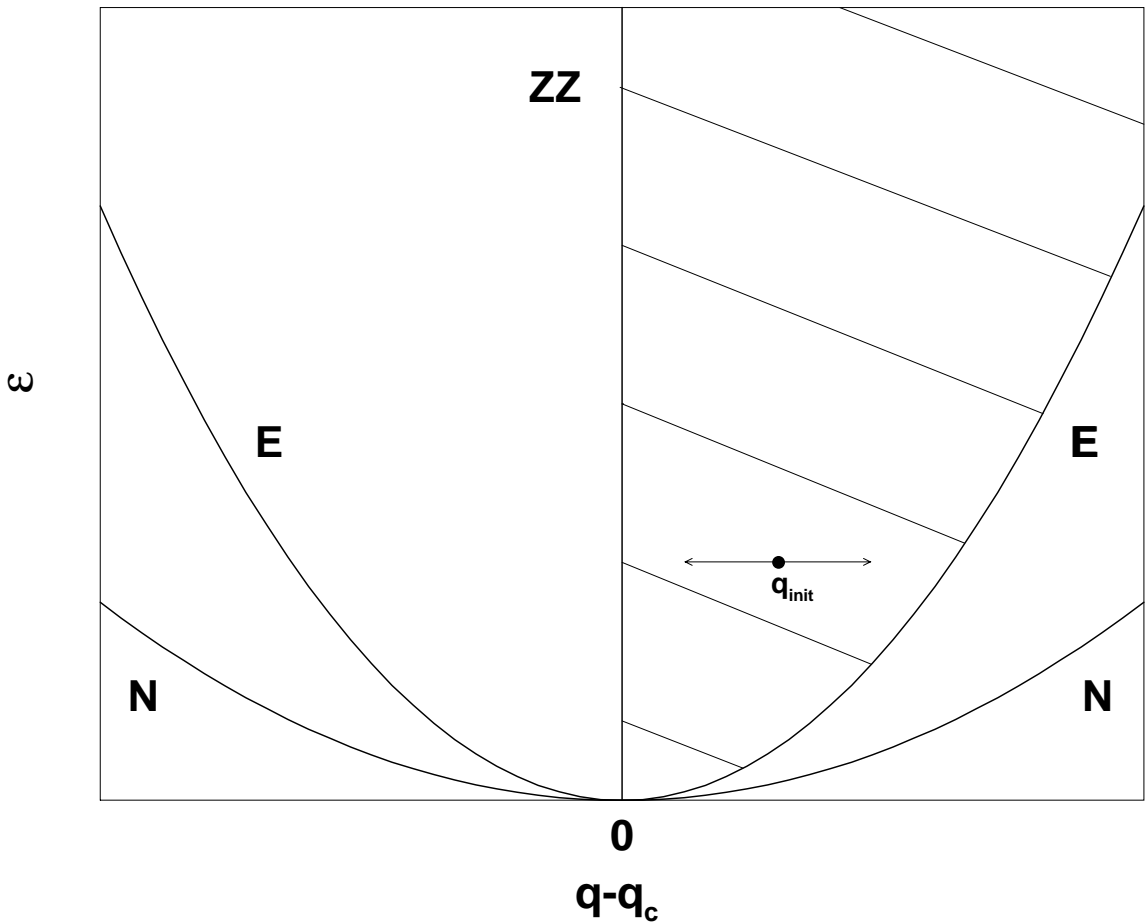,width=4.0in,angle=-90}}\
\caption{Schematic representation of the neutral stability curve (N),
the Eckhaus (E) and Zig-Zag (ZZ) boundaries for the
case of no flow. Within the shaded region the lamellar pattern is 
linearly stable. The effect of an imposed
shear on a uniform lamellar structure can be qualitatively understood
as the displacement of the state point along a line of constant $\epsilon$.
This representation, however, fails to give the correct location of the
stability boundaries given in the text.}
\label{fi:neutral_noshear}
\end{figure}

In order to gain some insight into the stability problem, we first
briefly review the known results for zero shear
\cite{re:kodama96,re:shiwa97}. In this case $A(t)$ is a
constant, and the matrix elements of $L(t)$ are independent
of time. An eigenvalue problem results by considering
solutions of Eq. (\ref{eq:lsa}) of the form 
$\delta A_\pm \sim e^{\sigma_+ t} + e^{\sigma_- t}$, and
instability follows when either eigenvalue is positive. Two modes of 
instability
are obtained: a zig-zag (ZZ) mode that leads to a transverse 
modulation of the lamellae (${\mathbf Q \cdot q } = 0$), and an
Eckhaus (E) mode that is purely longitudinal in nature ${\mathbf Q 
\cdot q } = Q q$. In the zig-zag case, $\sigma_+({\mathbf Q})$ has a 
maximum at 
\begin{equation}
Q_{max,ZZ}^2 = \frac{ 1 - 2 q^2 - 3 A^2}{2}.
\end{equation}
The eigenvalue $\sigma_+({\mathbf Q})$ changes sign on the line $q=q_c$,
which therefore defines the zigzag stability boundary.
In the Eckhaus case, we find after some straightforward algebra that
the perturbation with the largest growth rate is
\begin{equation}
Q_{max,E}^2 = \frac{ 64 \, \delta q^4 - (\epsilon - 
 4 \, \delta q^2)^2}{64 \, \delta q^2},
\end{equation}
with $\delta q=q - q_c$. Therefore the Eckhaus stability boundary is
given by $\epsilon = 12 \, \delta q^2$. These results are
schematically summarized
in Fig. \ref{fi:neutral_noshear}. The hatched area is the region of 
stability of a lamellar solution in the absence of shear flow. 
It is worth pointing out that this stability diagram is identical to
that of the Swift-Hohenberg model of Rayleigh-B\'enard convection
\cite{re:cross93}. Shiwa \cite{re:shiwa97} has recently shown that 
in the weak-segregation limit ($\epsilon \ll 1$), and in the absence 
of shear flow, the amplitude equation describing slow modulations of 
a lamellar solution is the same as the amplitude equation of the 
Swift-Hohenberg model near onset of convection. 
The same stability diagram has been derived by
Kodama and Doi \cite{re:kodama96} by examining free energy changes
upon distortion of a lamellar pattern.

We now return to the Floquet problem of Eq. (\ref{eq:lsa}) when
$A(t)$ is a periodic function of time (Eq. (\ref{eq:sol2})). Since
$A(t+T)=A(t)$ ($T = 2 \pi / \omega$), the solution of (\ref{eq:lsa}) 
is given by,
\begin{equation}
\left[ \begin{array}{c} \delta A_{+} \\ \delta A_{-} \end{array}
\right] = e^{\sigma t} \left[ \begin{array}{c} \phi_{+}(t) \\
\phi_{-}(t) \end{array} \right],
\end{equation}
with $\phi_{\pm}(t+T) = \phi_{\pm}(t)$. Equation (\ref{eq:lsa}) is
then transformed to an eigenvalue problem within $(0,T)$,
\begin{equation}
\label{eq:eigenvalue}
\frac{\partial}{\partial t} \left[ \begin{array}{c} \phi_{+}(t) \\
\phi_{-}(t) \end{array} \right] = - \sigma \left[ \begin{array}{c}
\phi_{+}(t)\\ \phi_{-}(t) \end{array} \right] + L(t) 
\left[ \begin{array}{c} \phi_{+}(t)\\ \phi_{-}(t) \end{array} \right].
\end{equation}

Given that the function $A(t)$ is quite complicated, we have solved
this eigenvalue problem numerically. The
eigenvalue $\sigma$ can depend in principle on the wavevector of the 
base state ${\mathbf q}$, on the wavevector of the perturbation
${\mathbf Q}$, and on the amplitude $\gamma$ and frequency $\omega$ of
the shear. For ease of presentation, we have focused on the case
$\epsilon = 0.04$ although extension to other values of $\epsilon$ is
straightforward. 

\begin{figure}[h]
\centerline{\psfig{figure=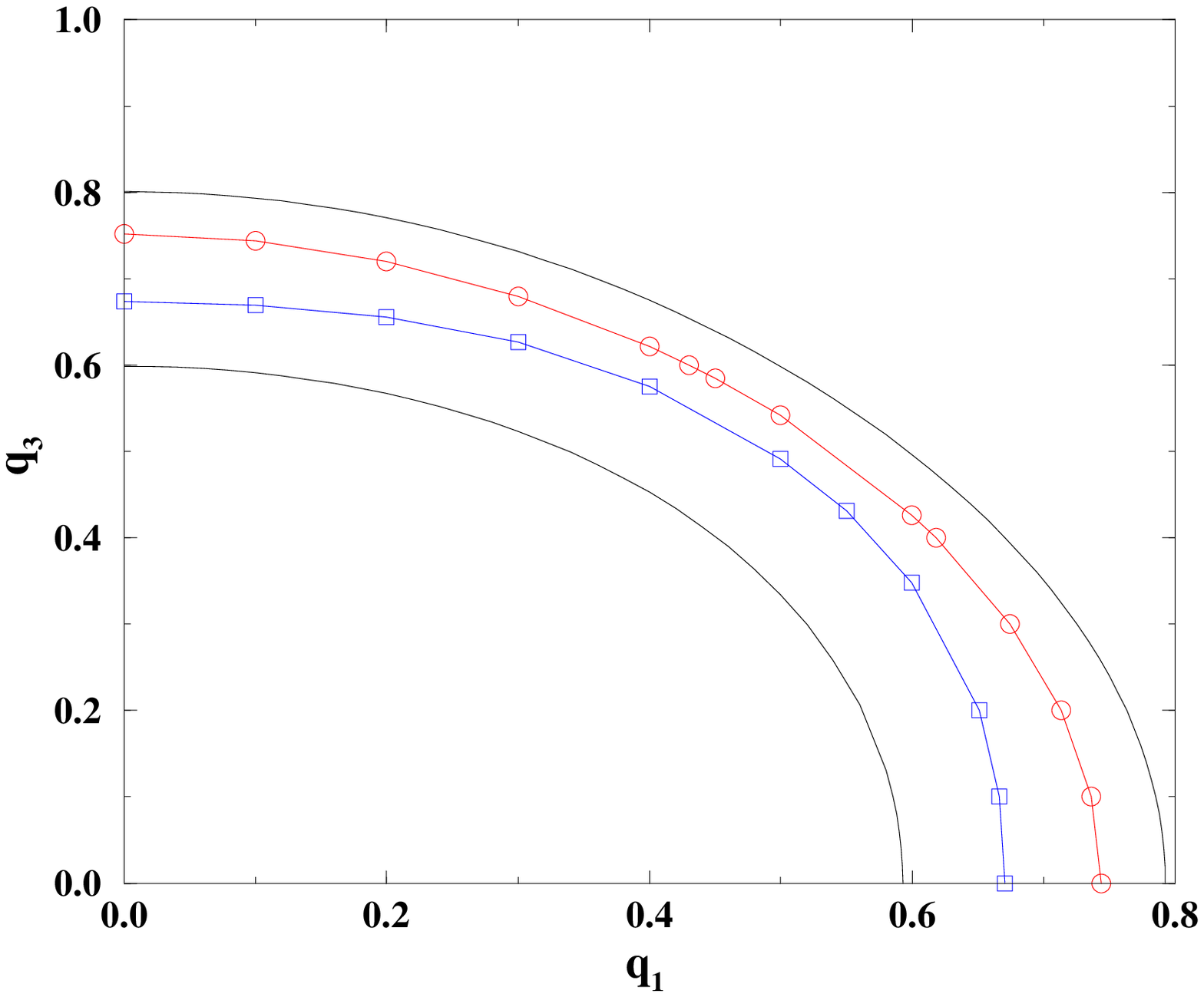,width=6in}}
\caption{Stability diagram of a lamellar pattern under oscillatory
shear of amplitude $\gamma = 0.2$ at $\epsilon = 0.04$. The solid
lines show the neutral stability curve. The two inner lines marked
with circles and squares bound the region of stability of the lamellar
structure against long wavelength perturbations.}
\label{fi:floqueta}
\end{figure}

\begin{figure}[h]
\centerline{\psfig{figure=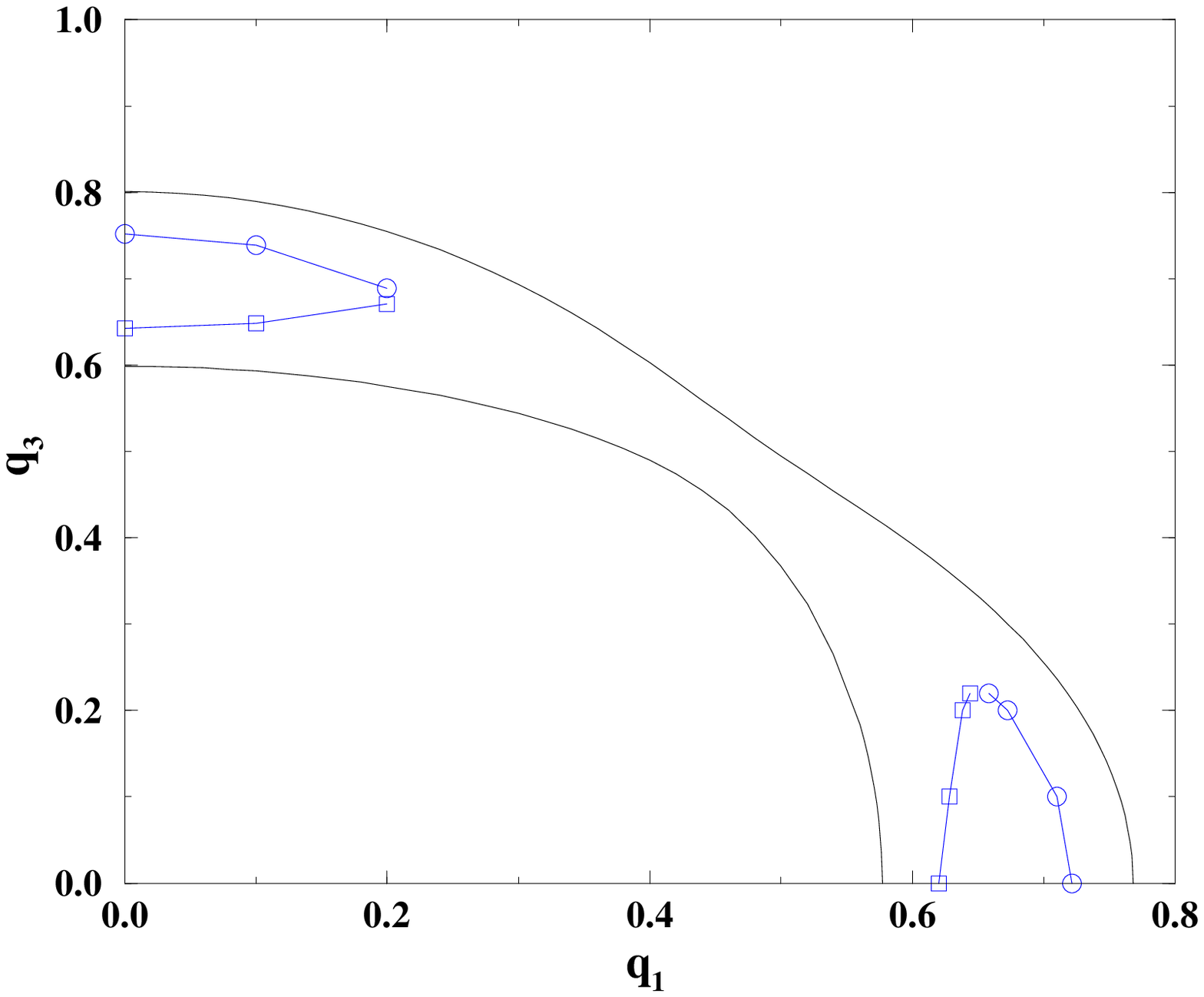,width=6in}}
\caption{Stability diagram of the lamellar pattern under oscillatory 
shear shear of amplitude $\gamma= 0.4$ at $\epsilon = 0.04$. The solid
lines show the neutral stability curve. The two inner lines marked
with circles and squares bound the region of stability of the lamellar
structure against long wavelength perturbations.}
\label{fi:floquetb}
\end{figure}

Figures \ref{fi:floqueta} and \ref{fi:floquetb}
summarize our results for the cases $\gamma =
0.2$ and $\gamma = 0.4$ respectively, and show the stability
boundaries in the plane $(q_{1},q_{3})$, as well
as the neutral stability curve already shown in
Fig. \ref{fi:neutral}. As before, $(q_{1},q_{3})$ is the wavevector of
the lamellar structure at $t=0$.
At fixed $\epsilon, \gamma$ and $\omega$, 
these curves have been obtained by determining the loci 
of ${\mathbf q}$ at which the
function $\sigma( {\mathbf Q})$ changes from a maximum to a saddle 
point at ${\mathbf Q}= 0$.
First we note that any orientation of the lamellar pattern that is not
initially close to either parallel or transverse is unstable to moderate
shears. Second, and up to our numerical accuracy, these curves
appear to be independent of angular frequency. Finally, and contrary 
to the case of no shear, the reciprocal basis vectors are not time
independent. Since the components of both ${\mathbf q}$ and ${\mathbf Q}$
are independent of time in the sheared frame, their mutual angle is
not (except for the case in which they are parallel). However, the
following statements can be made about the type of secondary instability.
We have found that the secondary instability is of the longitudinal 
type only when either $q_{1} = 0$ or $q_{3} = 0$ (intersections
between the lines marked with circles and the axes in Figs. 
\ref{fi:floqueta} and \ref{fi:floquetb}). Otherwise, the angle between
${\mathbf q}$ and ${\mathbf Q}$ is time-dependent. The lines on both
figures marked with squares have the property that even though
both ${\mathbf q}$ and ${\mathbf Q}$ are functions of time, their angle
oscillates periodically around $90^{o}$.


The cases discussed up to now concern long wavelength
instabilities of the base periodic pattern that are associated with
the broken translational symmetry of the original system by the 
appearance of a periodic pattern. We now show that it is possible to 
obtain analytical expressions for the stability boundaries against 
finite wavelength perturbations that may have some experimental 
relevance as well. In some experimental protocols, the 
lamellar pattern is first obtained in the absence
of shear. The resulting configuration
comprises regions or domains of locally parallel lamellae but with a
continuous distribution of orientations. A shear
flow is then initiated and the reorientation of the pattern studied as a 
function of time. The pattern obtained in the absence of shear may be now 
unstable to several finite wavenumber perturbations that would not
have been observable
in the case in which flow is present throughout the ordering
process. In the latter case
the unstable orientations would have decayed away during the
process of formation of the lamellae.  
In addition, the approximation that we derive below 
is generally valid when $Q_{3}$ cannot
approach zero, as is the case in a system of finite extent in the
direction of the velocity gradient.

We first define the following linear transformation,
\begin{equation}
\label{eq:lsa2}
 \left[ \begin{array}{c} {\delta_{+}} \\
 {\delta_{-}} \end{array}  \right]  =
  \left[ \begin{array}{cc} \frac{3 A(t)^2 l_-}{\sigma_+ - \sigma_-} &
\frac{\sigma_+ - \sigma_- + L_{22}- L_{11}}{2(\sigma_+ - \sigma_-)} \\
-\frac{3 A(t)^2  l_-}{\sigma_+-\sigma_-} & \frac{\sigma_+ - \sigma_- + L_{11}-
 L_{22}}{2(\sigma_+ - \sigma_-)} \end{array} \right]
\left[ \begin{array}{c}  \delta A_+ \\ \delta A_- \end{array} \right]
= T(t) \left[ \begin{array}{c}  \delta A_+ \\ \delta A_- \end{array} \right]
\end{equation}
which diagonalizes matrix $L(t)$, and where
\begin{equation}
\nonumber
\sigma_{\pm}(t) =  \frac{L_{11} + L_{22} \pm \sqrt{ (L_{11}-L_{22})^2 + 36 l_+
            l_- A(t)^{4}}}{2}.
\end{equation}
Combining Eqs. (\ref{eq:lsa}) and (\ref{eq:lsa2}), we find
\begin{equation}
\label{eq:dpm}
\frac{\partial}{\partial t} \left[ \begin{array}{c} {\delta_{+}(t)} \\
 {\delta_{-}(t)} \end{array}  \right]  =  \left[ \begin{array}{cc}
\sigma_+ & 0 \\ 0 & \sigma_- \end{array} \right]
\left[ \begin{array}{c}  \delta_+ \\ \delta_- \end{array} \right]
-   \frac{ A(t)^2 l_-}{\sigma_+ - \sigma_-}
\left[ \begin{array}{cc} \dot{M_1}  & - \dot{M_2}
\\ -\dot{M_1} & \dot{M_2} \end{array} \right]
\left[ \begin{array}{c}  \delta_+ \\ \delta_- \end{array} \right],
\end{equation}
where $\dot{M_1} = \frac{\partial}{\partial t} 
\left[ \frac{\sigma_+ - \sigma_- + L_{11} - L_{22}}{2 A^2 l_-} \right]$ and
$\dot{M_2} = \frac{\partial}{\partial t} \left[ \frac{\sigma_- - \sigma_+ 
+ L_{11} - L_{22}}{2 A^2 l_-} \right]$. 
For finite $Q$, $\frac{36 A(t)^4 l_+ l_-}{(L_{11}-L_{22})^2} \ll 1$.
Assuming $L_{11} - L_{22} > 0$ (the other case leads to no extra
complications), $\sigma_+ = L_{11}$ and $\sigma_- = L_{22}$.
Also $M_{1}=\frac{L_{11}-L_{22}}{A^2 l_-}$ and $M_2=0$ 
so that the equation for $\delta_+$ decouples from the equation for
$\delta_{-}$. The solution for $\delta_{+}$ is 
\begin{equation}
 \delta_+(t) = \delta_+ (0) e^{\int_{0}^{t}
(\sigma_+ - \frac{\partial}{\partial t'}  \ln \left[ \frac{L_{22}-L_{11}}
{A^2 l_-} \right]) dt'}. 
\end{equation}
The stability boundary is defined by $\bar{\sigma} = \int_{0}^{T}  
(\sigma_{+}  - \frac{\partial}{\partial t'}  \ln \left[ \frac{L_{22}-L_{11}}
{A^2 l_-} \right]) dt' =  \int_{0}^{T} \sigma_{+}   dt'=
0$. We have checked that this stability condition agrees with the
numerical stability analysis based on Eq. (\ref{eq:eigenvalue}) for
finite $Q$.

\begin{figure}[h]
 \hbox{\psfig{figure=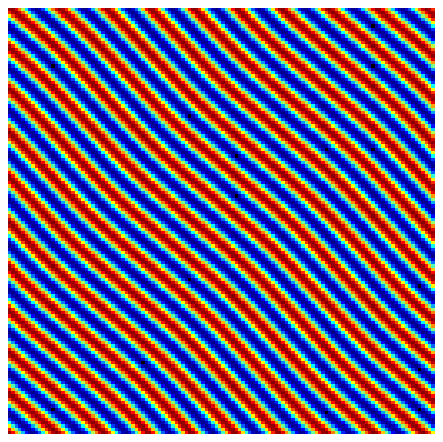,height=2.2in,width=2.2in}
   \psfig{figure=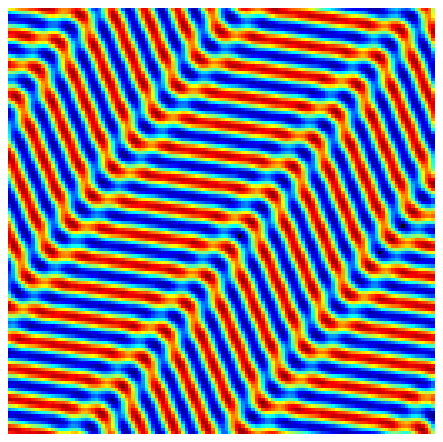,height=2.2in,width=2.2in}
   \psfig{figure=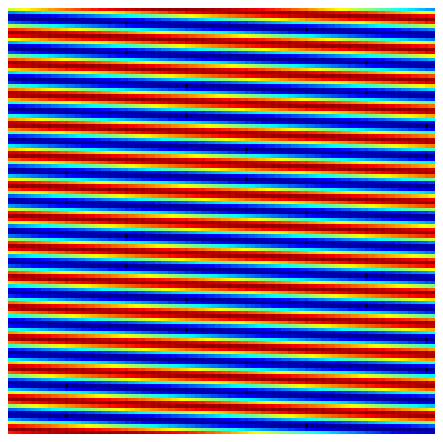,height=2.2in,width=2.2in} }
\vspace*{.5cm}

 \hbox{\psfig{figure=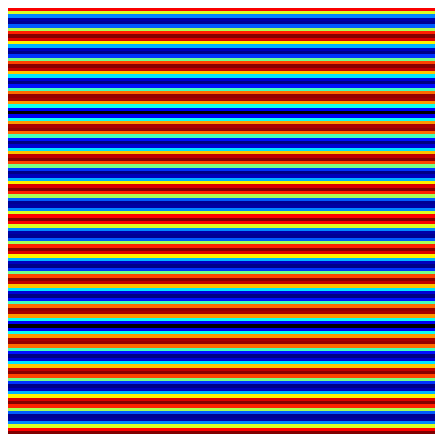,height=2.2in,width=2.2in}
   \psfig{figure=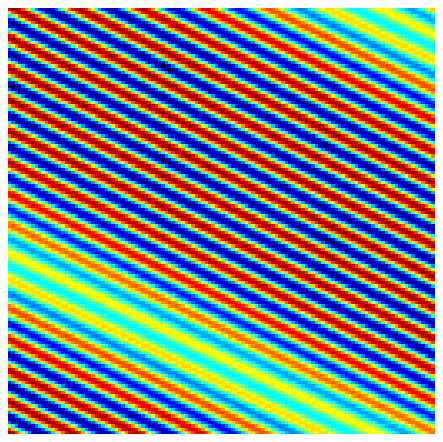,height=2.2in,width=2.2in}
   \psfig{figure=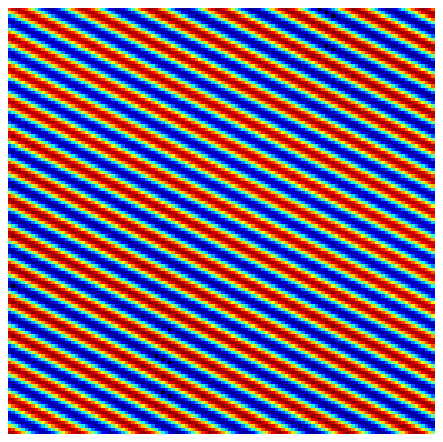,height=2.2in,width=2.2in} }
\vspace*{.5cm}


\caption{Results of the numerical integration of the model equation
to show the instability of a lamellar pattern (shown in grey scale). The
field $\psi$ shown in this figure has been transformed back to the 
laboratory frame of reference. Top, left to right: instability
approximately of the zig-zag type, followed by kink band formation
and reconnection leading to a different orientation. Bottom,
instability approximately of the Eckhaus type leading to a wavelength 
increase without change in the orientation.}
\label{fig:trans}
\end{figure}

We finish by illustrating the re-orientation dynamics of the lamellar
structure following a long wavelength instability by direct
numerical solution of the governing equation. We focus on the
region in which the uniform lamellar structure is linearly unstable.
The first example discussed concerns a lamellar structure of 
initial wavenumber $(q_1,q_3)=(-0.4908,0.4908)$ being sheared 
periodically with an amplitude $\gamma=1$.
Figures \ref{fig:trans}A,B,C show the
sequence of configurations obtained when $\omega = 5 \times 10^{-6}$. 
A long wavelength transverse 
modulation of the lamellae is observed (Figure \ref{fig:trans}A).
Subsequent growth leads to the formation of a forward kink band 
similar to that recently observed experimentally (Figure \ref{fig:trans}B)
\cite{re:polis96,re:polis98}. As the strain grows larger, 
the kink band disappears leaving behind a 
lamellar structure without any defects and oriented differently
relative to the shear (Figure \ref{fig:trans}C). 

In the second example (Figs. \ref{fig:trans}D,E,F), a structure initially
transverse to the flow is being sheared periodically with 
an amplitude $\gamma=1$ and a frequency $\omega = 5 \times 10^{-7}$.
A longitudinal perturbation is clearly visible that
manifests itself by local compression and dilation 
of the structure, leading to the disappearance of a pair of lamellae.
The overall result is an increase in the lamellar spacing 
without any change in the orientation. 

In summary, we have obtained a nonlinear solution of the model equations
that govern the formation of a lamellar structure in the 
weak-segregation limit. The solution is a periodic lamellar structure with
a time-dependent wavevector that adiabatically follows the imposed
shear flow, and a time-dependent amplitude which we have computed for
the cases of steady and oscillatory shears. In the case of an
oscillatory shear, the periodic solution only exists for a range of
orientations of the lamellae relative to the shear direction. The
width of the region depends on the shear amplitude but not on its
frequency. Long wavelength secondary instabilities further reduce the
range of existence of stable lamellar solutions. The corresponding
stability boundaries depend again on the shear amplitude, but are
independent (up to our numerical accuracy) of frequency. We next plan
to examine the stability of the nonlinear solution presented in this
paper when neither osmotic stresses nor viscosity contrast are neglected. 
 
\section*{Acknowledgments}
This research has been supported by the U.S. Department of Energy,
contract No. DE-FG05-95ER14566, and also in part by the Supercomputer
Computations Research Institute, which is partially funded by the
U.S. Department of Energy, contract No.  DE-FC05-85ER25000. F.D. is
supported by the Microgravity Science and Applications
Division of the NASA under contract No. NAG3-1885.

\newpage
\appendix
\label{app:algorithm}
\section{Numerical algorithm}

We use a pseudo-spectral technique to solve Eq. (\ref{eq:main}) in two
spatial dimensions and in the
sheared frame with periodic boundary conditions along both
directions. Equation (\ref{eq:main}) can be written as,
\begin{equation}
\label{eq:ft}
\frac{\partial \tilde\psi}{\partial t}= \sigma(t) \tilde\psi - q^2(t)
 \tilde{\psi^3},
\end{equation}
where $\sigma(t)= q^{2}(t) - q^{4}(t) - B$ and
$q^{2}(t)= q_1^2 + [a(t) q_1 - q_3]^{2}$.
The algorithm we use, due to Cross et al. \cite{re:cross94b}, 
is obtained by first multiplying both sides of Eq. (\ref{eq:ft}) by 
$\exp( -\sigma(t') \, t')$
and integrating over $t'$. This gives  
\begin{equation}
\exp(- \sigma(t) \, t') \tilde\psi \mid_t^{t+\Delta t} =  -q^{2}(t)
\int_t^{t+ \Delta t} dt' \tilde{\psi^3}(t') \exp(-\sigma(t) t'),
\end{equation}
where we have assumed $\sigma(t') \approx \sigma(t)$ and $q^{2}(t') \approx
q^{2} (t)$. Next,
we write the non-linear term $\tilde{\psi^3}(t')$ as a linear
function of $t'$
in the interval $t \leq t' \leq t+\Delta t$, i.e.,
\begin{equation}
\tilde{\psi^3}(t') \approx \tilde{\psi^3}(t) +
\frac{\tilde{\psi^3}(t+\Delta t)-\tilde{\psi^3}(t)}{\Delta t} (t-t').
\end{equation}
Combining the last two equations finally yields
\begin{eqnarray}
\label{eq:algo}
\nonumber
\tilde\psi(t+\Delta t) & = & \exp(\sigma(t) \, \Delta t) \tilde\psi(t)
- q^{2}(t)  \tilde{\psi^3}(t) \left[ \frac{\exp(\sigma(t) \, \Delta t)
 -1}{\sigma(t)}
\right] - \\ & & q^{2}(t) \left[ \frac{\tilde{\psi^3}(t+\Delta t)
-\tilde{\psi^3}(t)} {\Delta t} \right]
 \left[ \frac{\exp(\sigma(t) \, \Delta t) - (1 + \sigma(t) \Delta t)}
{\sigma^2(t)} \right].
\end{eqnarray}
Eq. (\ref{eq:algo}) is first evaluated with the last term on its
right-hand side set to zero. The resulting value for $\tilde\psi(t+\Delta t)$
is then used to estimate $\tilde{\psi^3}(t+\Delta t)$. Eq. (\ref{eq:algo})
is finally applied a second time  with all three terms on its
right-hand side now
included in the calculation. The fact that  the nonlinear terms
are integrated using an explicit procedure in time limits the  size of the
time step $\Delta t$ that can be used in simulations of the model.

All the numerical results
presented in this paper were obtained in the sheared frame of
reference with $128 \times 128$ 
spectral modes. We have chosen $B = 0.23$ (which
corresponds to $\epsilon = 0.04$) and a time step of maximum
size $\Delta t= 0.2$, for which no numerical instability was observed.
The initial condition $\psi ( {\mathbf r},t=0)$ is, unless otherwise
noted, a lamellar structure obtained by numerical integration of
Eq. (\ref{eq:algo}) with $a(t)=0$ (no shear) starting from a random
initial condition (a gaussian distribution for $\psi$ if zero mean and
small variance), for approximately 300,000 iterations until a
stationary lamellar structure is reached.

\bibliographystyle{achemso}
\bibliography{references}

\providecommand{\refin}[1]{\\ \textbf{Referenced in:} #1}
\begin{thebibliography}{10}

\bibitem{re:zchen97}
Chen,~Z.-R.;\ \ Issaian,~A.;\ \ Kornfield,~J.;\ \ Smith,~S.;\ \ Grothaus,~J.;\
  \ Satkowski,~M. \textit{Macromolecules} \textbf{1997,} \textsl{30,} 7096.

\bibitem{re:leibler80}
Leibler,~L. \textit{Macromolecules} \textbf{1980,} \textsl{13,} 1602.

\bibitem{re:ohta86}
Ohta,~T.;\ \ Kawasaki,~K. \textit{Macromolecules} \textbf{1986,} \textsl{19,}
  2621.

\bibitem{re:seul95}
Seul,~M.;\ \ Andelman,~D. \textit{Science} \textbf{1995,} \textsl{267,} 476.

\bibitem{re:bates90}
Bates,~F.;\ \ Fredrickson,~G. \textit{Ann. Rev. Phys. Chem.} \textbf{1990,}
  \textsl{41,} 525.

\bibitem{re:matsen94}
Matsen,~M.;\ \ Schick,~M. \textit{Phys. Rev. Lett.} \textbf{1994,} \textsl{72,}
  1994.

\bibitem{re:laradji97}
Laradji,~M.;\ \ Shi,~A.-C.;\ \ Noolandi,~J.;\ \ Desai,~R.
  \textit{Macromolecules} \textbf{1997,} \textsl{30,} 3242.

\bibitem{re:netz97}
Netz,~R.;\ \ Andelman,~D.;\ \ Schick,~M. \textit{Phys. Rev. Lett.}
  \textbf{1997,} \textsl{79,} 1058.

\bibitem{re:villain-guillot98}
Villain-Guillot,~S.;\ \ Netz,~R.;\ \ Andelman,~D.;\ \ Schick,~M.
  cond-mat/9803288.

\bibitem{re:fredrickson94}
Fredrickson,~G. \textit{J. Rheol} \textbf{1994,} \textsl{38,} 1045.

\bibitem{re:kodama96}
Kodama,~H.;\ \ Doi,~M. \textit{Macromolecules} \textbf{1996,} \textsl{29,}
  2652.

\bibitem{re:shiwa97}
Shiwa,~Y. \textit{Physics Letters A} \textbf{1997,} \textsl{228,} 279.

\bibitem{re:swift77}
Swift,~J.;\ \ Hohenberg,~P. \textit{Phys. Rev. A} \textbf{1977,} \textsl{15,}
  319.

\bibitem{re:greenside85}
Greenside,~H.;\ \ Cross,~M. \textit{Phys. Rev. A} \textbf{1985,} \textsl{31,}
  2492.

\bibitem{re:cross93}
Cross,~M.;\ \ Hohenberg,~P. \textit{Rev. Mod. Phys.} \textbf{1993,}
  \textsl{65,} 851.

\bibitem{re:elder92}
Elder,~K.;\ \ {Vi\~nals},~J.;\ \ Grant,~M. \textit{Phys. Rev. Lett.}
  \textbf{1992,} \textsl{68,} 3024.

\bibitem{re:cross95a}
Cross,~M.;\ \ Meiron,~D. \textit{Phys. Rev. Lett.} \textbf{1995,} \textsl{75,}
  2152.

\bibitem{re:gunton83}
Gunton,~J.~D.;\ \ {San Miguel},~M.;\ \ Sahni,~P.~S.  Kinetics of first order
  phase transitions.   In  , Vol.~8; Domb,~C.;\ \ Lebowitz,~J.,\ \ Eds.;
  Academic: London, 1983.

\bibitem{re:fredrickson88}
Fredrickson,~G.;\ \ Helfand,~E. \textit{J. Chem. Phys.} \textbf{1988,}
  \textsl{89,} 5890.

\bibitem{re:helfand89}
Helfand,~E.;\ \ Fredrickson,~G.~H. \textit{Phys. Rev. Lett.} \textbf{1989,}
  \textsl{62,} 2468.

\bibitem{re:gurtin96}
Gurtin,~M.~E.;\ \ Polignone,~D.;\ \ {Vi\~nals},~J. \textit{Math. Models and
  Methods in Appl. Sci.} \textbf{1996,} \textsl{6,} 815.

\bibitem{re:anderson98}
Anderson,~D.;\ \ McFadden,~G.;\ \ Wheeler,~A. \textit{Ann. Rev. Fluid Mech.}
  \textbf{1998,} \textsl{30,} 139.

\bibitem{re:cates89}
Cates,~M.;\ \ Milner,~S. \textit{Phys. Rev. Lett.} \textbf{1989,} \textsl{62,}
  1856.

\bibitem{re:iooss90}
Iooss,~G.;\ \ Joseph,~D. \textit{Elementary Stability and Bifurcation Theory;}
  Springer Verlag: New York, 1990.

\bibitem{re:polis96}
Polis,~D.;\ \ Winey,~K. \textit{Macromolecules} \textbf{1996,} \textsl{29,}
  8180.

\bibitem{re:polis98}
Polis,~D.;\ \ Winey,~K. \textit{Macromolecules} \textbf{1998,} \textsl{31,}
  3617.

\bibitem{re:cross94b}
Cross,~M.;\ \ Meiron,~D.;\ \ Tu,~Y. \textit{Chaos} \textbf{1994,} \textsl{4,}
  607.

\end{thebibliography}

\newpage

\newpage

\newpage

\newpage

\newpage

\end{document}